\documentstyle[preprint,prl,aps]{revtex}

\begin{document}
\draft
\title{Evidence for the singlet-dimer ground state in an $S$ = 1 antiferromagnetic bond alternating chain}
\author{Yasuo Narumi$^{1,2}$, Masayuki Hagiwara$^{3}$, Masanori Kohno$^{4}$, Koichi Kindo$^{2,1}$
}
\address{$^1$ CREST Japan Science and Technology Corporation (JST), Kawaguchi, Saitama 332-0012 Japan\\ 
$^2$ KYOKUGEN, Osaka University, Toyonaka, Osaka 560-8531 Japan\\ $^3$ RIKEN (The Institute of Physical and Chemical Research), Wako, Saitama 351-0198 Japan\\ $^4$ Mitsubishi Research Institute, Inc., Chiyoda, Tokyo 106-8141 Japan}
\date{Received }
\maketitle
\begin{abstract}
Susceptibility, ESR and magnetization measurements have been 
performed on single crystals of an $S$=1 bond alternating chain compound: [Ni(333-tet)($\mu$-NO$_2$)](ClO$_4$) (333-tet = ${\it N}$,${\it N}$'-bis(3-aminopropyl)propane-1,3-diamine) and the compound doped with a small amount of Zn. We observed an anomalous angular dependence in the Zn-doped sample. These behaviors are well explained by the model based on the VBS picture for the singlet-dimer phase. The picture implies that the free spins of $S$=1 with a positive single-ion anisotropy are induced at the edges of the chains without forming the singlet-dimer.
\end{abstract}

For the last two decades, studies of quantum spin systems have spread not only to linear-chain Heisenberg antiferromagnets (LCHA's) but also to LCHA's with complicated interactions, e.g., bond alternating interaction and next-nearst-neighbor interaction. The variety of the ground states appearing in these systems attracts much attention. Among them, the $S = 1$ antiferromagnetic bond alternating chain (BAC) has been the subject of great interest, because the system shows two different singlet ground states with an excitation gap, so-called singlet-dimer and Haldane, depending on the bond alternating ratio $\alpha$ between two different neighboring exchange constants, the excitation spectrum is gapless only at the phase separating point\cite{AF,SG,kato,yamamoto,totsuka,kitazawa,totsuka2}. These theoretical predictions have been checked by some numerical calculations and confirmed by recent experimental works (magnetization and susceptibility measurements)\cite{gapless,narumi}. 

The singlet-dimer and the Haldane states can be characterized distinctly by the representation based on the VBS(Valence-Bond-Solid) picture which was proposed by Affleck ${\it et} {\it al.}$ for the $S$=1 LCHA with a biquadratic term\cite{VBS}. According to this picture, spin-1 is obtained by symmetrization of two spin-1/2 variables. The Haldane state is constructed out of valence bonds connecting each neighboring pair of spin-1/2 as shown in Fig. 1(a). On the other hands, the siglet-dimer state consists of two valence bonds in coupled pair of spin-1 as shown in Fig. 1(b). We shall consider the situation of nonmagnetic impurity doping. In this case, the Haldane state becomes nearly fourfold degenerate because of two $S$ =1/2 degrees of freedom emerging at the both edges adjacent to the impurities as shown in Fig. 1(c). This interesting feature in the Haldane state has been firstly observed by the ESR experiment of Cu-doped NENP well known as a candidate compound for the $S$=1 LCHA\cite{hagiwara}. Applying this scenario to the singlet-dimer state, $S$ = 1 degrees of freedom are induced at the decoupled sites neighboring the impurities as shown in Fig. 1(d). It should be stressed that the single-ion anisotoropy is relevant only to the case of the $S$ = 1. In other words, the knowledge of the anisotoropy informs us whether the ground state is in the singlet-dimer phase or the Haldane phase. Therefore, the experimental study on the dilute system becomes a powerful method to investigate the ground state nature.  

In this paper, we report the results of susceptibility, ESR and magnetization measurements performed on single crystals of the BAC compound: [Ni(333-tet)($\mu$-NO$_2$)](ClO$_4$) (333-tet = ${\it N}$,${\it N}$'-bis(3-aminopropyl)propane-1,3-diamine) (abbreviated as NTENP) and the compound doped with a small amount of Zn (abbreviated as NTENP:Zn) . The experimental results are analyzed by the model based on the VBS picture and we will discuss the ground state of the BAC.

NTENP crystallizes in the triclinic system, space group $P\bar{1}$\cite{NO2}. The lattice constants and angles at room temperature are $a$=10.747(3)\AA, $b$=9.413(2)\AA, $c$=8.789(2)\AA, $\alpha$=95.52(2)$^{\circ}$, $\beta$=108.98(3)$^{\circ}$ and $\gamma$=106.83(3)$^{\circ}$. The chain is composed of Ni$^{2+}$ ions bridging through nitrito groups. The nitrito groups are disordered as is common in this kind of compound. Each chain is well isolated by ClO$_4^-$ counter anions through which no hydrogen bonds exist. The most important feature is that the inversion centers place not on the Ni$^{2+}$ ions but on the nitrito groups and two different bond distances of 2.142(3) and 2.432(6) \AA \ exist through them. As a result, this compound is expected to be the BAC. Escuer ${\it et} {\it al.}$ estimated the alternating ratio to be $\alpha$ = 0.88 by the susceptibility measurement of the polycrystalline sample\cite{NO2}.

Susceptibility data were collected with SQUID magnetometer
(Quantum Design's MPMS-XL7L) and magnetization measurements were performed by using a pulse magnet at KYOKUGEN in Osaka University. ESR measurements were carried out with a homemade spectrometer installed at RIKEN. We used the single crystals of pure and impure samples about 20 mg with an approximate dimension of 2mm $\times$ 2mm $\times$ 2mm for susceptibility and magnetization measurements. For the impure sample, a Zn concentration was estimated to be 3.5 $\pm$ 0.4 at.\% by a chemical analysis.   
Figure 2 shows the temperature dependence of the susceptibilities ($\chi(T)$) of NTENP and NTENP:Zn. The magnetic fields ($H$) are applied parallel and perpendicular to the chains for both samples. First we go into details about the pure NTENP. The $\chi(T)$'s have a round maximum at about 50 K for $H$ // chain and 53 K for $H \perp$ chain. The maximum value of the $\chi(T)$ for $H$ // chain is larger than that for $H \perp$ chain by about 10 \%. With decreasing temperature further, the $\chi(T)$'s of both axes fall gradually toward zero, showing that the compound has the singlet ground state with a finite spin gap. We calculated the $\chi(T)$ for $H$ // chain by Quantum Monte Carlo (QMC) method (96 sites). The Hamiltonian is defined as 
\begin{eqnarray}
{\cal H}_{chain} = J \sum_{i=1}^{L/2} (\mbox{\boldmath $S$}_{2i-1} \cdot \mbox{\boldmath $S$}_{2i} + \alpha \mbox{\boldmath $S$}_{2i} \cdot \mbox{\boldmath $S$}_{2i+1}) \nonumber \\
+ \sum_{i=1}^{L}D(S_i^z)^2 - \sum_{i=1}^{L}\mu_{\rm B}\mbox{\boldmath $S$}_i\tilde{g}\mbox{\boldmath $H$}, 
\end{eqnarray}
where $J$ is the nearest neighbor exchange constant for the strong
bond, $\alpha$ the bond alternating ratio, $L$ the size of the chain, $\mbox{\boldmath $S$}_i$ the $S$=1 spin operator at the site of $i$, $D$ the uniaxial single-ion anisotropy constant,  $S_i^z$ the $z$-component of $\mbox{\boldmath $S$}_i$, $\tilde{g}$ the $g$-tensor, $\mu_{\rm B}$ the Bohr magneton, and $\mbox{\boldmath $H$}$ the magnetic field. The chain direction is assumed to be the principal $z$-axis. Then the $\chi(T)$ along the chain is well reproduced by the parameters of $\alpha$ = 0.45, $J/k_B$ = 54.2 K, $g$ = 2.14 and $D/J$ = 0.25 ($D/k_B$ = 13.6 K) as shown in Fig. 3(a). Thus, the ground state of this compound must be in the singlet-dimer phase. The main reason for disagreement between our result and the $\chi(T)$ obtained by Escuer ${\it et } {\it al.}$ must be that the $\chi(T)$ of a polycrystal sample was analyzed without taking anisotropies of the single-ion and the $g$-values into account in the previous work.

Next we mention  the case of NTENP:Zn. Below about 50 K, the $\chi(T)$'s exhibit anomalous behaviors depending on the field direction, though the $\chi(T)$'s at high temperatures are quite similar to those of pure NTENP. For $H$ // chain, the $\chi(T)$ remains almost flat from 50 K to 5 K, and then drops suddenly toward zero as if a long range ordering or a spin-Peierls transition occurs. On the other hand, the $\chi(T)$ of $H \perp$ chain decreases slightly below about 50 K and rises markedly around 15 K with decreasing temperature further.  The coincidence of the $\chi(T)$'s of the impure sample with those of the pure one in the high temperature region shows that in spite of impurity doping the bulk nature of the BAC does not change so much. Let us discuss the ground state properties in view of the VBS picture. As we mentioned before, this compound must have the singlet-dimer ground state, therefore, $S$ =1 free spins are expected to be induced at the edges of the chain broken by Zn$^{2+}$ ions and to behave anisotropically because of the single-ion anisotropy. 

For a microscopic discussion of this picture, we performed ESR measurements of NTENP:Zn and show the frequency-field diagram for $H$ // chain (see Fig. 4). Observed ESR lines correspond to transitions between the singlet and the doublet states separated by the uniaxal single-ion anisotropy as shown in the inset of Fig. 4. It is noteworthy that the level crossing occurs at 3.5 T lower than the transition field originated from the spin gap of the bulk BAC as discussed later in  the magnetization measurements. According to the analysis based on the model of $S$=1 free spin with $D$-term, the Hamiltonian is given by 
\begin{equation}
{\cal H}_{edge} = D'(s^z)^2 - \mu_{\rm B}\mbox{\boldmath $s$}\tilde{g}'\mbox{\boldmath $H$},
\end{equation}
where \mbox{\boldmath $s$}, $D'$ and $\tilde{g}'$ are the spin operator, the uniaxal single-ion anisotoropy constant and $g$-tensor of the edge spin $s$, respectively. The ESR results gave us these parameters of $D'/k_B$ = 5.12 K and $g'_{//}$ = 2.17 for $H$ // chain. We notice that the $D'$ of the edge spin differs from the $D$ of the bulk chain. For $H \perp$ chain, we got the $g'_{\perp}$ = 2.13. The detail of the result will be published elsewhere.

Let us now return to the $\chi(T)$ analysis of NTENP:Zn. If the length $L$ of the chains broken by Zn$^{2+}$ is enough longer than the correlation length of the system, the edge spins may behave independently. Yamamoto calculated the spin correlation legth $\xi_{SS}$ of the BAC system and the singlet-dimer ground state with $\alpha$ = 0.45 has $\xi_{SS} \sim$ 5.8\cite{yamamoto}. In fact the chain length of this compound is estimated to be about 30 sites on the average according to the Zn concentration of $x$ = 0.035. We ignore the case that Zn$^{2+}$ ions are placed side by side because of the low concentration of Zn. Then, the $\chi(T)$ of NTENP:Zn $\chi_{imp}$ can be written as  
\begin{equation}
\chi_{imp} = (1-2x) \chi_{chain} + x \chi_{edge},
\end{equation}
where the $\chi_{chain}$ is the $\chi(T)$ of the bulk chain part and $\chi_{edge}$ is that of the edge part. Concretely speaking, we used the $\chi(T)$ of the pure sample as $\chi_{chain}$ and calculated exactly the $\chi(T)$ of the $S$ = 1 free spin with $D$-term as $\chi_{edge}$ using the parameters of $g$ and $D'$ obtained from the ESR measurement. Figure 3(b) shows the comparison between the results of the expetiment and the calculations with a fitting parameter $x$. The anomalous angular dependence can be  well explained by this superposition model for $x=0.033$. This $x$ value is reasonable for our chemical analysis.

In order to confirm the validity of this model, we have performed magnetization ($M(H)$) measurements. The experimental and calculated results are shown in Fig. 5 together with the $M(H)$ curves of the pure NTENP in the inset. For the pure NTENP, the $M(H)$ is almost zero below $H_c$ = 9.3 T and 12.4 T for $H$ // chain and $H$ $\perp$ chain, respectively ($H_c$: transition field). As the magnetic fields are applied further, we observe sharp increases with a convex curvature and then the $M(H)$'s increase almost linearly. It is obvious that these transitions are attributed to the level crossing between the singlet ground state and the excited triplet one. The $H_c$ for $H$ // chain is given by the following equation
\begin{equation}
H_{c//} = \frac{E_g-D/3}{g_{//}\mu_B}, 
\end{equation}
where $E_g$ is the energy gap of the BAC.
We obtained  $E_g/J$ = 0.33 ( $E_g/k_B$ = 17.9 K) showing that the ground sate is in the singlet-dimer phase with $\alpha \sim 0.43$\cite{kato,yamamoto,TNK}.

For $H$ // chain in the NTENP:Zn, we observe a sharp $M(H)$ bend around 3 T and a steep increase around 9 T. On the other hand, the $M(H)$ of $H \perp$ chain seems to be typical of a paramagnet below about 12 T, where the $M(H)$ shows a further increase. There is no doubt that these transitions at high fields come from the bulk chain of this compound. We emphasize that a sharp $M(H)$ bend around 3 T along the chain likely arises from the level crossing of the lowest states due to the edge spin as seen in the ESR experiment. Thus, we compare the $M(H)$ curves to the calculated curves with the same equation as in the $\chi(T)$ analysis. The $M(H)$ of NTENP:Zn $M_{imp}$ is given by
\begin{equation}
M_{imp} = (1-2x) M_{chain} + x M_{edge},
\end{equation}
where the $M_{chain}$ is the $M(H)$ of the pure NTENP and the $M_{edge}$ is the $M(H)$ calculated from the $S$ = 1 free spin with $D$-term. Here the same parameters as in the $\chi(T)$ analysis were used. We obtained a qualitatively good agreement, although the $M(H)$ increases above the transition fields are not very steep as those of the pure NTENP. The possible reason is that the system consists of assembly of the finite chains with different lengths. The gap energy decreases and converges a finite value with increasing the chain length\cite{kato}. Therefore the average transition field shifts to the high field side and the transition must become broad. Besides, we can not deny the possibility that the interaction between the edge spins and the bulk spins affects these magnetic properties to some extent.

In conclusion, we have performed susceptibility, ESR and magnetization measurements of single crystals of the bond alternating chain compound NTENP and NTENP doped with a small amount of Zn. These experimental results shows that this compound has the singlet-dimer ground state with $\alpha$ = 0.45 . Furthermore we have firstly observed the edge state of the bond alternating chain, which is well described by the VBS picture that the free spins of $S$ = 1 with a positive single-ion anisotropy are induced at the sites without forming the singlet-dimer.        

This study was partially supported by a Grant-in-Aid for Scientific Research from the Japanese Ministry of Education, Science, Sports and Culture, and by the MR Science Research Program of RIKEN. M. Hagiwara thanks to K. Katsumata for access to the ESR spectrometer installed at RIKEN. Thanks are also due to the Chemical Analysis Unit in RIKEN.

\begin{figure}[b,t,h]
 \vspace*{0cm}
 \caption{Schematic representations based on the VBS picture for the ground states of the $S$ = 1 linear chain systems: Haldane phase(a), singlet-dimer phase(b), nonmagnetic impurity doped Haldane phase(c) and nonmagnetic impurity doped singlet-dimer phase(d). The large open circles, the slashed circles and the small solid circles represent the atomic sites with $S$ = 1, the impurity sites and the spin-1/2 variables, respectively. The solid lines show the valence bonds.}
 \label{VBS}
\end{figure}

\begin{figure}
 \vspace*{0cm}
 \caption{Temperature dependence of the susceptibilities at 0.1 T on the single crystals of NTENP and NTENP:Zn.}
 \label{sus exp}
\end{figure} 
 
\begin{figure}
 \vspace*{0cm}
 \caption{Comparisons of the susceptibilities between the experiment and the calculation. The open circles in the upper panel (a) represent the observed susceptibility along the chain and the solid lines shows the numerical calculation by QMC (96 sites) with the parameter in the panel. The analysis of NTENP:Zn is shown in the lower panel(b). More details, see in the text.}
 \label{sus cal}
\end{figure}

\begin{figure}
 \vspace*{0cm}
 \caption{The frequency-field diagram of the ESR signals on the single crystal of NTPENP:Zn at 1.7 K in the magnetic field parallel to the chain.}
 \label{ESR}
\end{figure} 
 
\begin{figure}
 \vspace*{0cm}
 \caption{Magnetization curves of the single crystal of NTENP:Zn at 1.3 K and calculated magnetizatin curves. The inset shows the magnetization curves of the pure NTENP.}
 \label{mag}
\end{figure} 

\end{document}